\documentclass[12pt,reqno]{article}

\title{Self-Verifying Predicates in B\"uchi Arithmetic}

\author{Mazen Khodier\\
School of Computer Science\\
University of Waterloo\\
Waterloo, ON  N2L 3G1 \\
Canada\\
\href{mailto:mkhodier@uwaterloo.ca}{\tt mkhodier@uwaterloo.ca}\\
\and
Luke Schaeffer\\
Institute for Quantum Computing\\
University of Waterloo\\
Waterloo, ON  N2L 3G1 \\
Canada\\
\href{mailto:lschaeffer@uwaterloo.ca}{\tt lschaeffer@uwaterloo.ca}\\
\and Jeffrey Shallit\\
School of Computer Science\\
University of Waterloo\\
Waterloo, ON  N2L 3G1 \\
Canada\\
\href{mailto:shallit@uwaterloo.ca}{\tt shallit@uwaterloo.ca}}

\usepackage[usenames]{color}
\usepackage{amssymb}
\usepackage{amsmath}
\usepackage{amsthm}
\usepackage{amsfonts}
\usepackage{amscd}
\usepackage{graphicx}
\usepackage{booktabs}

\usepackage[colorlinks=true,
linkcolor=webgreen,
filecolor=webbrown,
citecolor=webgreen]{hyperref}

\definecolor{webgreen}{rgb}{0,.5,0}
\definecolor{webbrown}{rgb}{.6,0,0}

\usepackage{color}
\usepackage{fullpage}
\usepackage{float}

\usepackage{graphics}
\usepackage{latexsym}
\usepackage{epsf}

\def\AND{\, \wedge \, }
\def\OR{ \, \vee \, }
\def\union{\, \bigcup \,}
\def\suchthat{\ : \ }
\def\divides{\mid} 

\newcommand{\bfX}{{\bf X}}
\newcommand{\bfY}{{\bf Y}}
\DeclareMathOperator{\cnt}{Count}
\DeclareMathOperator{\per}{Per}
\DeclareMathOperator{\incr}{Incr}

\DeclareMathOperator{\eqfac}{EqFac}
\DeclareMathOperator{\eqrevfac}{EqRevFac}
\def\Enn{\mathbb{N}}

\makeatletter
\def\Ddots{\mathinner{\mkern1mu\raise\p@
\vbox{\kern7\p@\hbox{.}}\mkern2mu
\raise4\p@\hbox{.}\mkern2mu\raise7\p@\hbox{.}\mkern1mu}}
\makeatother

\begin{document}

\maketitle

\theoremstyle{plain}
\newtheorem{theorem}{Theorem}
\newtheorem{corollary}[theorem]{Corollary}
\newtheorem{lemma}[theorem]{Lemma}
\newtheorem{proposition}[theorem]{Proposition}

\theoremstyle{definition}
\newtheorem{definition}[theorem]{Definition}
\newtheorem{example}[theorem]{Example}
\newtheorem{conjecture}[theorem]{Conjecture}

\theoremstyle{remark}
\newtheorem{remark}[theorem]{Remark}

\begin{abstract}
We discuss a technique, based on Angluin's algorithm, for automatically generating finite automata for various kinds of useful first-order logic formulas in B\"uchi arithmetic. Construction in this way can be faster and use much less space than more direct methods.  We discuss the theory and we present some empirical data for the free software \texttt{Walnut}.
\end{abstract}

\section{Introduction}
\label{intro}
Let $k\geq 2$ be a fixed integer, and let $V_k (n) = \sup \{k^i \suchthat k^i \divides n\}$.
The first-order logical theory
$\langle \Enn, +, <, 0, 1, V_k \rangle$, is
sometimes called {\it B\"uchi arithmetic}
\cite{Buchi:1960}.    This theory, an 
extension of Presburger arithmetic, is powerful enough to express finite automata,
and is algorithmically decidable \cite{Bruyere&Hansel&Michaux&Villemaire:1994}.
Notice that addition and subtraction of integer variables are permitted in this theory, but not multiplication or division (although one can
express multiplication and integer division by
fixed integer constants).

In particular, this theory is quite useful in combinatorics on words, since it can be used to decide many claims about automatic sequences, provided the underlying numeration system is addable \cite{Shallit:2023}. Here by an addable numeration system we mean one for which a finite automaton can compute the addition relation $x+y = z$.
We say a sequence $(a(n))_{n \geq 0}$ is {\it automatic\/} if it takes its values in a finite set, and there is a finite automaton that, on input $n$ expressed in the numeration system $\cal N$, 
reaches a state with output $a(n)$; see \cite{Allouche&Shallit:2003}.

\begin{example} It is decidable, given an automatic sequence ${\bf a} = (a(n))_{n \geq 0}$ defined on an addable numeration system,
whether $\bf a$ is {\it squarefree}; that is, whether it contains no two consecutive identical nonempty blocks.

This follows because the property of not having a square is first-order expressible, as follows:
$$ \neg\exists i,n \ (n\geq 1) \AND \forall t \ (t<n) \implies a(i+t) = a(i+n+t) .$$
\end{example}

More precisely, the fundamental theorem of B\"uchi arithmetic is the following:  
\begin{theorem}
There exists an algorithm that takes a first-order logical formula $\varphi$ about an automatic sequence in an addable numeration system, phrased in terms of addition and indexing of integer variables, and computes an automaton that accepts exactly those values of the free (unbound) variables that make $\varphi$ true.  If there are no free variables, the computed automaton is a single state that accepts everything {\tt (TRUE)} or rejects everything {\tt (FALSE)}.
\label{buchi}
\end{theorem}

However, the worst-case running time of the decision procedure is truly formidable; it is of the form
$$2^{2^{\Ddots^{ 2^{p(N)}}}},$$
where the number of $2$'s in the exponent is equal to the number
of quantifier alternations in the formula, $p$ is a polynomial, and $N$ is the size
of the logical formula.
This is because the algorithm depends on repeated applications of the subset construction from automata theory
\cite{Hopcroft&Ullman:1979}, each one of which can potentially result in an exponential blowup in the number of states.

Integers in B\"uchi arithmetic are represented by words $x$ over a finite alphabet $\Sigma^*$.  We also need to represent $j$-tuples of integers $(n_1, \ldots, n_j)$; this is done by using the alphabet $\Sigma^j$, and then $n_i$ is the integer whose representation is
the projection of the $i$'th coordinate
of $x$.  This may require padding of the
shorter inputs with leading zeros.  If
$x \in (\Sigma^j)^*$, we let
$[x]$ denote the tuple of integers
represented by $x$.   The canonical
representation of an integer $n \in \Enn$
(that is, the one without leading zeros) is written $(n)$.  This is generalized to $j$-tuples by
writing $(n_1, n_2, \ldots, n_j)$.  Throughout the paper, we assume representations of integers are read with the most significant digit first.

The paper is organized as follows.  In Section~\ref{walnut} we discuss the {\tt Walnut} program and what it does.  In Section~\ref{angluin} we review Angluin's algorithm.  The next sections cover how we can use the self-verifying property of predicates to create an automaton for equality of factors (Section \ref{equality}), equality of reversals of factors (Section \ref{reversals}), periods (Section~\ref{periods}), recognizing the addition relation (Section \ref{adder}), and summation (Section~\ref{summation}).  Section~\ref{algeng} discusses algorithm engineering issues and reports some computational results for our very preliminary implementation.

\section{ {\tt Walnut} }
\label{walnut}

An automaton-based decision procedure for
B\"uchi arithmetic, as well as certain related theories based on other kinds of representations, has been implemented in the
free software \texttt{Walnut} \cite{Mousavi:2016}.  It has been used in over a hundred research papers and books to confirm old results, correct mistakes in the literature, and prove entirely new results
\cite{Shallit:2023}.

The decision procedure for B\"uchi arithmetic implemented 
in \texttt{Walnut}
allows representing integers in a variety of addable numeration systems, including base $k$ for integers $k\geq 2$, Fibonacci (Zeckendorf) numeration system, Tribonacci numeration system, and so forth. 
The user can also define their own numeration system.

Given a first-order logical formula $\varphi$,
{\tt Walnut} produces a DFA 
(deterministic finite automaton) accepting the values of the free variables that make 
$\varphi$ true.  Furthermore, we are guaranteed that the automata that
{\tt Walnut} computes are minimal.  {\tt Walnut} can also produce DFAO's (deterministic finite automata with output), where the output is a specified function of the last state reached.

In some cases, {\tt Walnut} returns no response to a query because its limit on space is exceeded (roughly, that automata cannot have more than $2^{32}$ transitions.)  This can be the case even if the final result would be small, because intermediate results can be extremely large.  This can occur even with a single quantifier and a very simple formula.   

As an example of the kinds of space consumption that can occur, consider the following first-order logical formula for equality of factors of an infinite word $\bfX$:
$$ \eqfac(i,j,n) := \forall t \ (t<n) \implies \bfX[i+t]=\bfX[j+t] .$$
It is true precisely when the length-$n$ factor beginning at position $i$ is the same as that beginning at position $j$.  
(In this predicate, as in all others we discuss in this paper, the domain of integer variables is assumed to be $\Enn = \{ 0,1,2,\ldots \}$,
the natural numbers.)   Since {\tt Walnut} implements B\"uchi arithmetic, it can form an automaton evaluating this logical formula for automatic sequences $\bfX$.\footnote{The need for B\"uchi arithmetic, as opposed to the simpler Presburger arithmetic, arises because we need to be able to express the $i$'th term of an automatic sequence
${\bf X}[i]$.} For example, for the Thue-Morse sequence, the resulting automaton takes the inputs $i, j, n$ in parallel, has $14$ states, and the largest intermediate automaton formed during the computation has $408$ states.  To construct it, one can use the \texttt{Walnut} command
\begin{verbatim}
def tm_eqfac "At (t<n) => T[i+t]=T[j+t]":
\end{verbatim}
However, when we carry out the same construction for the so-called Tribonacci word \cite{Barcucci&Belanger&Brlek:2004,Tan&Wen:2007} ${\bf tr} = 010201001020101020100102\cdots ,$
using the \texttt{Walnut} command
\begin{verbatim}
def trib_eqfac "?msd_trib At (t<n) => TR[i+t]=TR[j+t]":
\end{verbatim}
the largest intermediate automaton has 323,831,403 states (!), while the final result has only $26$ states.\footnote{For this particular example, some reformulations of the logical predicate result in much smaller intermediate automata.}

Therefore it is desirable to have alternative methods to compute some of these basic and useful automata.
In this paper we develop  a new technique for this, of both theoretical and practical interest.  It uses the observation that some of the most useful predicates are
``self-verifying'' in a certain sense.  In some cases, we can prove that our new algorithm runs in time polynomial in the size of the automaton computing the original sequence $\bfX$ and the size of the final result.   In particular, the idea is based on Angluin's algorithm, discussed in the next section.

The idea of using self-verifying predicates to verify automata was already discussed in a number of papers, such as
\cite{Rampersad&Shallit:2025,Shallit:2024}.  The novelty of this paper is that no ``guessing'' of the automaton to be verified is needed; the automaton is, so to speak, ``built from the ground up'' in a completely deterministic fashion.

\section{Angluin's algorithm}
\label{angluin}

Dana Angluin developed an algorithm
\cite{Angluin:1987}, sometimes called the ``$L^*$ algorithm'', that allows a {\it learner\/} to infer a finite automaton for a regular language $L \subseteq \Sigma^*$ from examples and counterexamples.   In her algorithm, the learner interacts with a {\it teacher\/} in two different ways.   

The learner can present a word $w \in \Sigma^*$ to the teacher and ask if $w \in L$; this is called a {\it membership query}.  The learner can also present an hypothesized automaton $A$ for the language $L$ and ask if $L(A) = L$.   If the teacher answers positively, the learner has now found an automaton for $L$ and the algorithm terminates.   If, on the other hand, the teacher answers negatively, they provide a counterexample to the learner; that is, a member of the symmetric difference 
$(L \setminus L(A)) \union (L(A) \setminus L)$.
We call all of this a {\it hypothesis query}.
If the minimal automaton for $L$ has $N$ states, then the total number of membership queries and hypothesis queries, and their size, is bounded by a polynomial in $N$.   More precisely, her result is the following:
\begin{theorem}[\cite{Angluin:1987}]
    \label{thm:angluin}
    There exists an algorithm $L^{*}$ that learns a regular language $L$ by performing membership and hypothesis queries. If all negative hypothesis queries return counterexamples of length at most $m$ and the regular language $L$ has a minimal DFA of  $n$ states, then the total running time is polynomial in $m$ and $n$.
\end{theorem}

In Section~\ref{optimizing} we will need a few more details about how Angluin's algorithm works to infer $L \subseteq \Sigma^*$.  It maintains a 
finite prefix-closed set of strings $S$, a finite
suffix-closed set of strings $E$, and a map $T$ from 
 $(S \union S\Sigma)E$ to $\{0,1\}$, where the intent is that $1$ means the string is in $L$ and $0$ otherwise.  Prefix-closed means every prefix of $S$ is in $S$, and similarly for the suffix-closed property of $E$.  Thinking of $T$ as a table with rows
 labeled by $S \union S\Sigma$ and columns labeled
 by $E$, the table $T$ is said to be closed if for each $t \in S\Sigma$ there is an $s \in S$ such that
 the row labeled $s$ equals the row labeled $t$.  It is called consistent if whenever two rows are equal, labeled $s_1$ and $s_2$, say, then the row labeled
 $s_1 a$ equals that labeled by $s_2 a$, for all $a \in \Sigma$.

Normally the teacher and the learner are separate entities.  However, for certain kinds of languages, corresponding to certain logical formulas about automatic sequences, we can use Theorem~\ref{buchi} to construct an algorithm that plays the role of {\it both\/} learner and teacher, as we will see in the next section.   Currently we have no general theory characterizing exactly for which kinds of logical formulas our technique works.  Nevertheless, the general idea is widely applicable and adaptable to a number of situations, as we intend to show in the following sections.

\section{Equality of factors}
\label{equality}

One of the most useful of all predicates for understanding the properties of an infinite word $\bfX$ is $\eqfac$, the {\it equality of factors} predicate we saw in Section~\ref{intro}; that is, given integers $i,j,n$, determine whether the length-$n$ factors beginning at positions $i$ and $j$ are the same.  
If we want to understand aspects of $\bfX$ such as factor complexity (aka subword complexity), the number of distinct blocks of length $n$ appearing in $\bfX$, then finding an automaton for this $\eqfac$ is a critical building block.

However, because of the appearance of the $\forall$ quantifier, the best estimate we can find for the complexity of finding the automaton using Theorem~\ref{buchi}, without more detailed analysis, is an exponential upper bound on the size of the resulting automaton.   We can potentially see exponential blowup in intermediate automata, even if the final result is small.

In this section, by combining Angluin's algorithm with a self-verifying predicate, we prove the following result:
\begin{theorem}
If $\bfX$ is an automatic sequence, we can construct a minimal automaton $A$ for the $\eqfac$ predicate in time polynomial in the size of the automaton for $\bfX$ and the size of $A$.
\label{thm3}
\end{theorem}

The language $L$ in Angluin's algorithm
is 
$$L = \{ x \in {\cal V}_3 \suchthat
[x] = (i,j,n) \text{ and }
\eqfac(i,j,n) \},$$
where ${\cal V}_j$ is the set of all valid representations of
integer $j$-tuples in the
underlying numeration system.  For example, if we are representing integers in base $2$, then
${\cal V}_3 = (\{0,1\}^3)^*$.
In particular, we note that a standard convention is that if an input is accepted, then so are all inputs that start with an arbitrary number of leading zeros.  We enforce this by insisting that from the initial state, a transition on the input $[0,0,\ldots, 0]$
always leads back to the initial state.

\subsection{Membership queries}

We now explain how the membership queries for $L$ required by Angluin's algorithm can be carried out in polynomial time for $\eqfac$.  Of course, by ``polynomial time'' we mean polynomial time in the {\it number of bits\/} of the integers involved.

Given $x$, the first thing we check is that we have used only legal representations in the underlying numeration system.
For example, if we are working in the Fibonacci numeration system, we have to check that the projection of $x$ into the binary strings representing three integers $i_0,j_0,n_0$ have no two consecutive $1$'s.  If any of them do have two consecutive $1$'s, then $x \not\in L$.

Now we know that the input is a legal representation of a triple of integers $(i_0,j_0,n_0)$.  We want to determine if
$\bfX[i_0..i_0+n_0-1]=\bfX[j_0..j_0+n_0-1]$.
A small difficulty is that given an automaton for $\bfX$ that maps
$n$ to $\bfX[n]$, it is not clear how many states are needed for
$(i,n) \rightarrow \bfX[i+n]$.  This is because the
obvious logical formulation of this is
$\exists t\ t=i+n \AND \bfX[t]$, and the
$\exists$ quantifier could create an NFA, which is then determinized.  Thus, at least
theoretically, this could introduce
exponential blowup.  We can avoid this using De Morgan's law to rewrite a $\forall$ query to a $\exists$ formula.  Furthermore, we avoid computing the DFA for $\exists x \ \varphi(x)$; instead we compute the automaton for $\varphi(x)$ and use breadth-first search (BFS) to determine whether any string is accepted, and if so, what a shortest such string is.

Thus, the first step is create the automaton $E(i,j,n,t,u,v)$
for the expression
$$ (t<n) \AND (u=i+t) \AND (v=j+t) \AND (\bfX[u] \neq \bfX[v]).$$
The automata for the first three conjuncts each have a constant number of states, and so do the automata for the intersection of the corresponding languages.  The last conjunct can be expressed as an automaton using $O(N^2)$ states, where $N$ is the
number of states in the automaton for $\bfX$.
Constructing the automaton for $E(i,j,n,t,u,v)$ needs to be done only once, at the
very beginning of the algorithm.

Now we want to evaluate $\eqfac(i_0,j_0,n_0)$ for some specific
$(i_0, j_0, n_0)$.   To do this,
we create automata accepting
$0^*(i_0)$, $0^*(j_0)$, and $0^*(n_0)$.
These individually require only $O(\log\max(i_0,j_0,n_0))$ states.
With additional intersections we create an automaton $E'$ accepting only those $6$-tuples $(i,j,n,t,u,v)$
for which $i$ matches $i_0$, $j$ matches
$j_0$, and $n$ matches $n_0$.  We now perform
breadth first search in $E'$, to see if $E'$ accepts anything; that is, if an accepting state is reachable from the start state.
This can be done in linear time in the size of $E'$, which is polynomial in $\log\max(i_0,j_0,n_0)$.   If $E'$
accepts anything, then $\eqfac(i_0, j_0, n_0)$ is false; otherwise it is true.

Thus we can perform membership queries in polynomial time.

\subsection{Hypothesis queries}
\label{hyp_quer}

Now we explain how to implement hypothesis queries efficiently.
Given an automaton $A$, we want to check whether $L(A) = L$.   The crucial point is that $\eqfac$ is \textit{self-verifying}; that is, provided certain logical formulas about $A$ hold, they provide us with a proof by mathematical induction that the automaton $A$ is correct.  In the case of $\eqfac$, an easy induction on $n$ shows that a putative automaton $A$ correctly decides
$\eqfac$ if and only if all of 
the following assertions hold:
\begin{enumerate}
    \item $A$ accepts no illegal representations
    for $i,j,n$ (representations that are not
    permissible in the given
    numeration system);
    \item the initial state of $A$ transits to
    itself on input $[0,0,0]$;
    \item $\forall i,j \ A[i,j,0]$;
    \item $\forall i,j,n\  
        A[i,j,n+1] \iff (A[i,j,n] \AND \bfX[i+n]=\bfX[j+n])$.
\end{enumerate}
Conditions 3 and 4 form the basis of a proof by induction on $n$ that $A$ is correct.
Although it is perhaps not immediately obvious, we can check all of these conditions in polynomial time; in the case of conditions 3 and 4, by reformulating them with De Morgan's laws. 

Condition 3 can be restated as
$$ \neg \exists i,j,t \ (t=0) \AND (\neg A[i,j,t]) .$$
This can be checked by complementing $A$,
intersecting with the automaton accepting 
$(\Sigma^*, \Sigma^*, 0^*)$, and using BFS to check if there is a path to an accepting state. If there is none, then
condition 3 holds; otherwise it fails
and a counterexample $(i_0, j_0, 0)$ is given by the label of the path to any accepting state.

For Condition 4, we first rewrite it as
\begin{align*}
& \forall i,j,n,t,u,v \ (t=n+1 \AND u=i+n \AND v=j+n) \implies \\
& \quad (A[i,j,t] \iff (A[i,j,n] \AND \bfX[u]=\bfX[v])) 
\end{align*}
and then use de Morgan's law
to replace the $\forall$ with the conjunction of the following three statements:
\begin{align*}
& \neg \exists i,j,n,t \ (t=n+1) \AND (\neg A[i,j,n]) \AND A[i,j,t] \\
& \neg \exists i,j,n,t,u,v\ (t=n+1) \AND (u=i+n) \AND (v=j+n) \AND (\neg A[i,j,t]) \AND A[i,j,n] \\
& \AND \bfX[u]=\bfX[v] \\
& \neg \exists i,j,n,t,u,v \ (t=n+1) \AND (u=i+n) \AND (v = j+n) \AND \bfX[u]\neq \bfX[v] \AND A[i,j,t] .
\end{align*}
In all three cases we can test the conditions by forming intersections of languages created by the direct product of automata, followed by BFS to check if an input is accepted.  If no input is accepted, then $A$ is the correct automaton.  Otherwise, $A$ accepts some word that provides the needed counterexample $(i_0, j_0, n_0)$.  Actually, we do not know, a priori, whether it is $A[i_0, j_0, n_0]$ or
$A[i_0,j_0,n_0+1]$ that is wrong, so some additional membership
queries may be necessary to determine
which value is incorrect.

We can now put everything together to prove
Theorem~\ref{thm3}.

\begin{proof}
As the analysis of Angluin's algorithm
shows, if $A$ has $t$ states then we 
do at most polynomially many membership queries, and the construction of our automata for the membership queries shows that the size of the queries is also polynomially bounded.  Furthermore, at most polynomially many hypotheses about $A$ are evaluated, so the total cost is polynomial time.
\end{proof}

Similarly, we can check formulas like
$$ \forall t \ (t<n) \implies \bfX[i+t]=\bfY[j+t]$$
for two automatic sequences $\bfX$ and
$\bf Y$ defined over the same numeration
system.

In what follows, we omit explicit mention of conditions 1 and 2 above, although we always need them to check that a hypothesized automaton $A$ is correct.

\section{Equality of reversals of factors}
\label{reversals}

Another self-verifying predicate involves checking equality of a reversal of a factor.
Let $w^R$ denote the reversal of the word
$w$ so that, for example,
$({\tt drawer})^R = {\tt reward}$.
Another useful predicate, $\eqrevfac$, asserts that
$\bfX[i..i+n-1]=\bfX[j..j+n-1]^R$.   As a special
case, we can use this predicate to test whether a factor is a palindrome.

We can use exactly the same technique as in the previous section, based on the fact that $\eqrevfac$ is self-verifying just the way $\eqfac$ is.  More precisely,
if $A$ is a claimed automaton for
$\eqrevfac$, it is correct if and only
if 
\begin{enumerate}
    \item $\forall i, j \ A[i,j,0]$; 
    \item $\forall i,j, n \ A[i,j,n+1] \iff (A[i+1,j,n] \AND \bfX[i]=\bfX[j+n])$.
\end{enumerate}
We leave the details to the reader.

\section{Periods of factors}
\label{periods}

Let $x = x[1..n]$ be a finite word.
We say that $p$ is a {\it period} of
$x$ if $x[i]=x[i+p]$ for all $i$
such that $1 \leq i \leq n-p$.  From the definition we see that $0$ is trivially a period of every word.  Furthermore, if $x$ is of length $n$,
then every integer $\geq n$ is also trivially a period of $x$.  Often these edge cases are not considered valid periods, but in order to show that period is self-verifying, we need to regard them as periods.

Consider the predicate
$\per(i,n,p)$ that asserts that $p$ is a period of $\bfX[i..i+n-1]$.
This is a very basic assertion that can be used to determine whether
a given factor is an overlap 
(of the form $axaxa$, where $a$ is a single letter and $x$ is possibly empty), a square (of the form $xx$ with $x$ nonempty), a cube
(of the form $xxx$ with $x$ nonempty), and so forth, and it is very desirable to be able to find the corresponding automaton efficiently.  We can write a first-order logic formula for this as follows:
$$\forall t\ (t+p<n) \implies \bfX[i+t]=\bfX[i+t+p],$$
and use Theorem~\ref{buchi} to find
an automaton computing it.  However, this may result in a large intermediate automaton, even when the final result is small.

Of course, we can also express $\per(i,n,p)$ using the $\eqfac$ predicate, but it is possible that the automaton for $\per$ is smaller and hence easier to compute directly by our method.

Suppose $A$ is an automaton that is claimed to compute the predicate
$\per(i,n,p)$ that asserts that $p$ is a period of $\bfX[i..i+n-1]$.  Then $A$ is correct if and only if the
following conditions hold.
\begin{enumerate}
    \item $\forall i,p \ A[i,0,p]$;
    \item $\forall i,n,p \ A[i,n+1,p]
    \iff ((p\geq n+1) \OR (p \leq n \AND A[i,n,p] \AND \bfX[i+n]=\bfX[i+n-p])) $.
\end{enumerate}
These can be translated into $\exists$ predicates just as we did for the case of $\eqfac$.  Notice that the subtraction can be handled by introducing a new variable $t$ and imposing the condition
$t+p=i+n$.

\section{Creating an adder for a numeration system}
\label{adder}

Using \texttt{Walnut} requires choice of a numeration system.  For some numeration systems, such as base $k$ for $k \geq 2$, the numeration system is built in.  

One of the crucial requirements for \texttt{Walnut}'s algorithm to succeed is that the numeration system be \textit{addable}; that is, there is a finite automaton recognizing the addition relation $x +y = z$.

For more exotic numeration systems, such as those based on a Pisot number \cite{Frougny:1992,Frougny&Solomyak:1996,Bruyere&Hansel:1997}, constructing the adder can be a bit of an onerous task.  However, Angluin's algorithm allows it to be constructed ``from the ground up'' when it exists,
because the adder itself is self-verifying.
This was already pointed out in \cite[Remark 2.1]{Mousavi&Schaeffer&Shallit:2016}, and we simply reprise this below.

The first step is that we need an incrementer; that is, an automaton $\incr$ that accepts a pair of inputs $(n,x)$ in parallel if and only if $x = n+1$.
This can be generated by combining the automaton recognizing the legal representations in the numeration system with an automaton that compares two inputs
lexicographically, as discussed in
\cite[p.~37]{Schaeffer:2013}.  

Membership queries are more or less trivial and do not require \texttt{Walnut}.

Now suppose we have an automaton $A$ that we hypothesize computes the addition relation.
We claim $A$ is correct if and only if both of the following conditions hold (in addition to the conditions on
the validity of the representations and leading zeros).
\begin{enumerate}
    \item $\forall y,z\ A[0,y,z] \iff y=z$;
    \item $\forall x,y,z,t,u\ (\incr(x,t) \AND \incr(z,u)) \implies (A[x,y,z] \iff A[t,y,u])$.
\end{enumerate}
These two conditions provide the basis and the induction step, respectively, for an induction proof on $x$ that $A$ is correct.  

These two conditions can be rearranged, as we did above, to consist of existential claims that can be verified efficiently with BFS.  If they fail, short
counterexamples can be computed with BFS, too.

For other recent discussions of ``automatically'' obtaining an adder for a numeration system,
see \cite[\S 4.1]{Mignoty&Renard&Rigo&Whiteland:2024} and
\cite{Carton&Couvreur&Delacourt&Ollinger:2025}.

\section{Summation of synchronized sequences}
\label{summation}

We say a sequence (or function) $(b(n))_{n \geq 0}$ from
$\Enn$ to $\Enn$ is {\it synchronized\/} if there
exists an automaton $B$ that accepts the representation of those pairs
$(n,x) \in \Enn\times \Enn$ for which $x = b(n)$.
In this case we call $B$ synchronized also.
See \cite{Shallit:2021} for more information about these sequences.  Notice that it is possible that $n$ is represented in one numeration system, while $x$ is represented in some other numeration system.  For an example of this, see \cite{Shallit:2024}, where $n$ is represented in base $4$ while $x$ is represented in base $3$.

Often we would like to do partial summation on
such a sequence, defining
$c(n) = \sum_{0 \leq i < n} b(i)$.  Unfortunately there are examples where $a$ is synchronized but the partial sum sequence $b$ is not.  As an example, consider the synchronized
sequence ${\bf p}_2 = 11010001 \cdots$, which is $1$ if $n+1$ is a power of $2$ and $0$ otherwise.
Then the partial sum sequence of ${\bf p}_2$
is given by $
 \lfloor \log_2 n \rfloor + 1$ for $n\geq 1$.  However, this kind of growth rate is impossible for a
synchronized sequence \cite{Shallit:2021}.

Nevertheless, in some cases the partial sum sequence is synchronized, and if it is, then we would like to construct the automaton for it.
Luckily, the assertion that $c$ is the partial
sum sequence for $b$ is self-verifying.  In addition to the usual checks involving legal representations and leading zeros, in order to verify that an automaton $C$ for $c$ is correct,
we only need to check that $C$ satisfies the following:
\begin{itemize}
    \item[(a)] $\forall x \ C[0,x] \iff x=0$;
    \item[(b)] $\forall n,t,u,y,z \ 
    (u=n+1 \AND z=t+y \AND B[n,t]) \implies
    (C[n,y] \iff C[u,z])$.
\end{itemize}
Condition (a) provides the basis, and condition (b) the induction step for a proof by induction on $n$ that 
$C[n,x]$ holds if and only if $c(n) = x$.  These two conditions can be turned into existential claims that can be verified with BFS, as before.

There is also the issue of how to compute membership queries.  Here we are presented with $n$ and $z$ and we want to determine whether
$c(n) = z$.  
This can be done as follows:  from the synchronized automaton $B$ for $b$, first determine a linear representation for $b(n)$.  Recall that a linear representation consists of a row vector $v$, a column vector $w$, and a matrix-valued morphism $\gamma$ such that
$b(n) = v \cdot \gamma(y) \cdot w$ for all
representations $y$ of $n$ (including those with leading zeros).  This linear representation can be trivially computed in linear time in the size of $B$, as explained in \cite[\S 9.8]{Shallit:2023}.  

Once we have the linear representation for $b$, we can compute a linear representation for $c$
using a fairly simple transformation that only increases the size of the linear representation by a constant factor \cite{Shallit&Stipulanti:2025}.  And from the linear representation we can compute $c(n)$ and check to see whether it equals $x$, in polynomial time in $\log n$.

Thus once again we can use Angluin's algorithm to compute the automaton $C$ for the partial sum sequence $c$, if it exists.  Unfortunately, we have no general method to know {\it a priori} whether $C$ exists.  If it does not, Angluin's algorithm will run forever.  If it does halt, however, we are guaranteed that the computed automaton $C$ is correct and that the running time is bounded by a polynomial in the size of $B$ and the size of a minimal automaton for $C$.

In some cases \texttt{Walnut} can handle those sequences $b$ that take negative integer values.
For example, this is true for $k$-automatic sequences, which we can accept using an automaton defined over base $-k$, and for Fibonacci-automatic sequences, which we can accept using an automaton in the negaFibonacci system. In some cases, then we can compute automata for partial sums $\sum_{0 \leq i < n} b(i)$
for {\it integer\/} sequences, not just sequences of natural numbers. For more details about negative numbers in {\tt Walnut}, see
\cite{Shallit&Shan&Yang:2023}.

One of the most useful examples of this technique is 
the symbol-counting predicate 
$\cnt_a (n,x)$ which is true if and only if $x = |\bfX[0..n-1]|_a$,
the number of occurrences of the symbol $a$ in the length-$n$ prefix of $\bfX$.  
  If Angluin's algorithm produces an automaton for $\cnt_a (n,x)$, 
then we can also compute $|\bfX[i..i+j-1]|_a$ via subtraction.  This  is essential in understanding the abelian properties of an automatic sequence, such as the presence or absence of abelian powers.  (Recall that a factor $x$
is an abelian $k$'th power if
$x = x_1 x_2 \cdots x_k$, with each
$x_i$ a permutation of $x_1$.)

We implemented Angluin's algorithm to successfully deduce the automaton for the partial sums of the Thue-Morse word, Fibonacci word, and Tribonacci word; see Table~\ref{runtimes2} for a summary of the computation.  We also successfully deduced the rarefied sum automaton from
\cite{Shallit:2024}, which accepts $n$ represented in base $4$ and $\sum_{0 \leq i < n} (-1)^{t(3i)}$ represented in base $3$, where $t(i)$ is the $i$'th bit of the Thue-Morse sequence.

\section{Algorithm engineering issues}
\label{algeng}

\subsection{Membership queries}
In practice (as opposed to theory), there are various algorithmic engineering techniques to speed up the approach we have outlined in this paper.
For example, for the equality of factors predicate, membership queries involve testing whether
$\bfX[i..i+n-1] = \bfX[j..j+n-1]$
for some specific triple $(i,j,n) = (i_0, j_0, n_0)$.
There are three ways to evaluate this,
as follows:

First, we can evaluate this predicate directly from the automaton for $\bfX$.   
In general, this uses 
$\Theta(n)$ operations on integers of magnitude
$\Theta(\max(i,j,n))$, which is clearly not polynomial time in $\log n$.  Nevertheless, in practice this is likely to be extremely fast, 
so if $n_0$ is small (say $n_0<10^7$) then we can evaluate the query more efficiently in this way.

Second, we can evaluate this predicate by
substituting the specific values $(i,j,n) = (i_0, j_0, n_0)$ in the predicate and evaluating the resulting predicate in
\texttt{Walnut}.  This could end up taking even more time, asymptotically, since it involves creating intermediate
automata computing the
functions $t \rightarrow \bfX[i_0 + t]$ and
$t \rightarrow \bfX[j_0+t]$, which, when
combined by $\bfX[i_0 + t]= \bfX[j_0+t]$, could theoretically result in an automaton of size $\Theta(\max(i_0,j_0)^2)$.  There is also another practical limitation here:  in its current implementation, \texttt{Walnut} cannot process queries involving integers larger than $2^{31} -1$.   
So it seems that this approach is unlikely to be competitive in practice with the first approach, except perhaps if $i_0, j_0$ are small (say $<10^3$) and $n_0$ is large (say $10^7 < n_0 < 2^{31}$).

The third approach, previously outlined in Section~\ref{hyp_quer}, uses \texttt{Walnut} to build the “equality of factors” automaton without quantifiers once, and then intersects it with small automata that accept only the specific values of the tuples we wish to test.  This can be done in polynomial time.

Similarly, computing partial sums $ \sum_{0\leq i < n} b(i)$ of a synchronized sequence $(b(i))$ can be done either directly in $O(n)$ time, or via linear representations
 with $\log n$ multiplications
of a vector times an $r \times r$ matrix, where $r$ is the rank of the linear representation.  This costs
$\Theta(r^2 \log n)$.
For small $n$ the direct method will beat the linear
representation, so some software engineering is needed
to make sure membership queries are as efficient as
possible.

Another way to compute the linear representation for the partial summation is as follows.  If $b(n,x)$ is the synchronized automaton for $b$, then the {\tt Walnut} command
\begin{verbatim}
def sumb "Ex $b(n,x) & i<n & j<x":
\end{verbatim}
will directly compute the linear representation
(as a {\tt Maple} file) for $c(n) = \sum_{0 \leq i < n} b(i)$.  Unlike the method of the previous paragraph, because of the presence of the $\exists$ quantifier, it is conceivable that this could result in exponential blowup (in the size of the automaton for $b$).

\subsection{Optimizing performance}
\label{optimizing}

A primary bottleneck in our implementation of Angluin's algorithm is the consistency test. Aside from the fact that every table entry needs to be queried once to be filled, the check for consistency requires querying strings that are not even in the table. In the worst case, to check for consistency we must perform $\Theta(|S| \cdot \Sigma \cdot |E|)$ membership queries. However, many membership queries could actually be looking at the same string but split at different points into $S$ and $E$. Thus, we cached query results into a hash table. This simple optimization technique reduced the total runtime of the algorithm on the equality of factors predicate for the Thue-Morse sequence from 1672.2 s to 124.6 s. Other runtime statistics, after using a hash table, are shown for different words in Table \ref{runtimes}. The execution times in the table are measured in seconds.

\begin{table}[ht]
  \centering
  \caption{Performance of the $\eqfac$ predicate on four automatic sequences.}
  \scriptsize
  \setlength{\tabcolsep}{4pt}
    \centering
    \begin{tabular*}{0.85\textwidth}{@{\extracolsep{\fill}} l c c c c @{}}
      \toprule
      Metric       & Thue-Morse & Baum-Sweet & Fibonacci & Tribonacci  \\
      \midrule
      Substitution method time &  89.8  & 4402  &  323.1   &  2046.4   \\
      Intersection method time  &  439.5  & 20230.3     & 296.1    &  2687.1    \\
      \# of unique queries       &  1672  & 75243 &  1032    &  4816   \\
      \# of incorrect hypotheses    &    7    & 43    &      6   &      11   \\
      Longest counterexample   &    4    & 8     &      3   &       7   \\
      Longest queried string  &    8    & 15    &      6   &      11    \\
      Final \# of states      &   15    & 130   &     12   &      27   \\
      Final $|S|$                   &   26    & 210   &     16   &     40    \\
      Final $|E|$                   &    9    & 51    &      9   &     17    \\
      \bottomrule
    \end{tabular*}
  \label{runtimes}
\end{table}

\begin{table}[ht]
  \centering
  \caption{Performance of the partial and rarefied sum predicates on different automatic sequences.}
  \scriptsize
  \setlength{\tabcolsep}{4pt}
  \begin{minipage}[t]{0.68\textwidth}
    \centering
    \begin{tabular*}{\textwidth}{@{\extracolsep{\fill}} l c c c @{}}
      \toprule
      \multicolumn{4}{c}{Partial sums}\\
      \midrule
      Metric                       & Thue-Morse     & Fibonacci     & Tribonacci     \\
      \midrule
      Total time (in seconds)                        &  1.4  &  32.2   &  3243.4  \\
      \# of unique queries       & 132  & 146 & 12932        \\
      \# of incorrect hypotheses                       &     3   &      3   &      23  \\
      Longest counterexample                       &     3   &      4   &       11  \\

      Longest queried string                       &     6   &      7   &      18  \\
      Final \# of states                       &    7   &     7   &      89  \\
      Final $|S|$                       &    8   &     11   &      133  \\
      Final $|E|$                        &     5   &      4   &      32  \\
      \bottomrule
    \end{tabular*}
  \end{minipage}
\begin{minipage}[t]{0.16\textwidth}
  \centering
  \begin{tabular*}{\textwidth}{@{\extracolsep{\fill}} c @{}}
    \toprule
    \multicolumn{1}{c}{Rarefied sums \cite{Shallit:2024}} \\
    \midrule
      Thue-Morse  \\
    \midrule
      6156.2 \\
      3548 \\
      9 \\
      4 \\
      9 \\
      17 \\ 
      29 \\
      11 \\
    \bottomrule
  \end{tabular*}
\end{minipage}
\label{runtimes2}
\end{table}


\subsection{Walnut considerations}

Several engineering details related to Walnut, and outside the core algorithm, can affect performance. First, reusing a single \texttt{Walnut} process throughout the run avoids repeated startup overhead. Second, we found \texttt{Walnut}'s built-in BFS in the \texttt{test} command to be inefficient. So, instead we used Python implemented a simple BFS for finding shortest counterexamples. Also, as mentioned in Section \ref{hyp_quer}, to ensure each membership test remains polynomial-time, we rewrote key predicates in order to avoid \texttt{Walnut} running into exponential behavior.

We also faced a problem when trying to reject invalid representations. For example, we need to use \texttt{?msd\_fib} to let \texttt{Walnut} know that we are currently using the Fibonacci numeration system. However, in the case of the Fibonacci word if we write all of the counterexample predicates using 
\texttt{?msd\_fib}, we won't actually be able to reject the invalid representations. This is due to the fact that \texttt{Walnut} automatically does not accept these invalid representations, so they will never be suggested as counterexamples as they are not part of the test. To work around that, we added one predicate at the beginning for rejecting invalid representations and used \texttt{?msd\_2} instead. This technique is needed no matter what numeration system we are dealing with. The reason is that 
\texttt{Walnut} is able to consider all binary sequences when using \texttt{?msd\_2}, so we are able to isolate the invalid ones in this separate predicate.

Finally, a crucial consideration is the ordering of the counterexample predicates. Since the hypothesis tests are essentially doing an induction, we must ensure that the already accepted strings are in fact correct. Otherwise, we will be enforcing the algorithm to accept even more incorrect strings instead of rejecting them. For instance, the last hypothesis query mentioned in \ref{hyp_quer}, was not written as a single test when programmed. Instead, each conjunction was used as a separate test and their orders were reversed. By using the third conjunction first, we ensure that we reject incorrect strings before building upon them as in the second conjunction.

\bibliographystyle{splncs04}
\bibliography{self}

\end{document}